\documentclass[apjl]{emulateapj}

\usepackage{natbib}
\usepackage{ifpdf}

\pdfoutput=1

\usepackage[bf, sf, FIGTOPCAP, nooneline, tight]{subfigure}
\usepackage{graphicx}
\usepackage{dcolumn}
\usepackage{bm}
\usepackage[usenames,dvipsnames]{color}
\usepackage[colorlinks=true, linkcolor=BrickRed, citecolor=Blue, urlcolor=Blue, filecolor=Blue]{hyperref}
\usepackage{booktabs}
\usepackage{hhline}

\def\apj{ApJ}%
\def\apjs{ApJS}%
\def\mnras{MNRAS}%
\def\prd{Phys.~Rev.~D}%

\def\healpix{HEALPix~}
\def\deg{\ifmmode^\circ\else$^\circ$\fi}
\def\doubleline{\vskip 3pt\hrule \vskip 1.5pt \hrule \vskip 5pt}

\begin{document}

\title{Power asymmetry in {\it WMAP} and {\it Planck} temperature sky maps \\
  as measured by a local variance estimator}

\author{Y. Akrami\altaffilmark{1}, Y. Fantaye\altaffilmark{1,2}, A. Shafieloo\altaffilmark{3,4}, H. K. Eriksen\altaffilmark{1}, F. K. Hansen\altaffilmark{1}, A. J. Banday\altaffilmark{5,6}, and K. M. G\'{o}rski\altaffilmark{7,8}}

\altaffiltext{1}{Institute of Theoretical Astrophysics, University of Oslo, P.O. Box 1029 Blindern, N-0315 Oslo, Norway}
\altaffiltext{2}{Department of Mathematics, University of Rome Tor Vergata, Rome, Italy}
\altaffiltext{3}{Asia Pacific Center for Theoretical Physics, Pohang, Gyeongbuk 790-784, Korea}
\altaffiltext{4}{Department of Physics, POSTECH, Pohang, Gyeongbuk 790-784, Korea}
\altaffiltext{5}{Universit\'{e} de Toulouse, UPS-OMP, IRAP, F-31028 Toulouse cedex 4, France}
\altaffiltext{6}{CNRS, IRAP, 9 Av. Colonel Roche, BP 44346, F-31028 Toulouse cedex 4, France}
\altaffiltext{7}{Jet Propulsion Laboratory, California Institute of Technology, 4800 Oak Grove Drive, Pasadena, CA, USA}
\altaffiltext{8}{Warsaw University Observatory, Aleje Ujazdowskie 4, 00-478 Warszawa, Poland}

\email{yashar.akrami@astro.uio.no}
\email{y.t.fantaye@astro.uio.no}
\email{arman@apctp.org}

\begin{abstract}
We revisit the question of hemispherical power asymmetry in the {\it WMAP}
and {\it Planck} temperature sky maps by measuring the local variance over
the sky and on disks of various sizes. For the 2013 {\it Planck} sky map
we find that none of the 1000 available isotropic {\it Planck} ``Full Focal
Plane'' simulations have a larger variance asymmetry than that
estimated from the data, suggesting the presence of an anisotropic
signature formally significant at least at the $3.3\sigma$ level. For
the {\it WMAP} 9-year data we find that 5 out of 1000 simulations have a
larger asymmetry. The preferred direction for the asymmetry from the
{\it Planck} data is $(l,b)=(212^\circ,-13^\circ)$, in good agreement with
previous reports of the same hemispherical power asymmetry.

\end{abstract}


\section{Introduction}

The assumptions of statistical isotropy and homogeneity of the
Universe on very large scales, jointly called the cosmological
principles, are two of the main pillars of our present standard model
of cosmology. In the past, the validity of these assumptions was
based largely on philosophical arguments, or as a necessity in order
to simplify otherwise complicated equations. Today, the situation is
very different. With the advent of advanced space and ground-based
instruments, stringent tests of these basic assumptions are
available. 

Indeed, almost immediately after the first release of the measurements
of the cosmic microwave background (CMB) temperature fluctuations by the {\it Wilkinson
Microwave Anisotropy Probe} ({\it WMAP}) experiment \citep{bennett:2003},
different groups found various anomalous features in the data that
hinted at possible violations of the statistical isotropy \citep[see, e.g.,][]{Tegmark:2003ve,deOliveiraCosta:2003pu,Vielva:2003et,Larson:2004vm,Land:2004bs,Land:2005ad,McEwen:2004sv,McEwen:2006yc,Jaffe:2006sq,Hinshaw:2006ia,Spergel:2006hy,Cruz:2006fy,Bridges:2006mt,Copi:2006tu,Land:2006bn,Bernui:2005pz,Bernui:2008ve,Pietrobon:2008rf}. Among
these was one suggesting a directional dependency of the CMB angular
power spectrum \citep{eriksen2004,hansen2004}, often referred to as a
``hemispherical power asymmetry''. Since then, this observation has
been probed with different methods and algorithms, and is with current
data sets generally found to be statistically significant at the
$3\sigma-3.5\sigma$ level depending on the algorithm and angular scales
under considerations
\citep{minkowski,Eriksen:2007pc,hansen2009,hoftuft2009,Axelsson:2013mva,2013arXiv1303.5083P}.
Accordingly, theorists are now attempting to reconcile these
observations with the current cosmological standard model
\citep[see, e.g.,][]{Erickcek:2008sm,Dai:2013kfa,Lyth:2013vha,Liu:2013kea,Liddle:2013czu,Mazumdar:2013yta,Namjoo:2013fka,Abolhasani:2013vaa,Cai:2013gma,Kohri:2013kqa,McDonald:2013qca,Kanno:2013ohv}.

\citet{gordon:2005} suggested that the power asymmetry might be
modeled in terms of a dipole modulation of the form
\begin{equation}
\frac{\Delta T}{T}|_{\textrm{mod}}(\hat{n})=(1+A \,\hat{n}\cdot\hat{p})\frac{\Delta T}{T}|_{\textrm{iso}}(\hat{n})\label{eq:dipole},
\end{equation}
where $\frac{\Delta T}{T}|_{\textrm{iso}}$ and $\frac{\Delta T}{T}|_{\textrm{mod}}$ are,
respectively, the isotropic and modulated CMB temperature fluctuations
along a direction $\hat{n}$ on the sky, $A$ is the amplitude of the
dipole modulation and $\hat{p}$ is the preferred direction. Direct
likelihood fits of this particular model have been reported by
\citet{Eriksen:2007pc,hoftuft2009,2013arXiv1303.5083P} for the {\it WMAP} and {\it Planck} \citep{2013arXiv1303.5062P}
data, obtaining a typical dipole modulation amplitude of $A\sim0.07$
on large angular scales, statistically significant at
$\sim3\sigma$. Equivalent results have been obtained using for
instance the BiPolar Spherical Harmonics (BiPoSHs) technique
\citep{Hajian:2003qq,Hajian:2006ud,2013arXiv1303.5083P}. On small angular scales, the
dipole amplitude is much lower, and appears to vanish by a multipole
moment of $\ell\sim500$--600 \citep{hansen2009,2013arXiv1303.5083P}.

However, there is a significant debate in the field concerning whether
these findings are statistically significant after accounting for
so-called {\it look-elsewhere} effects,\footnote{The look-elsewhere effect is a statistical effect that impacts the calculated significance of observing a local excess of events when searching for a signal in a possible range of a particular quantity without knowing a priori where the signal will appear within the range. This is especially severe if the significance is moderate. The significance calculation must account for the fact that an excess anywhere in the range could equally be considered as a signal. Therefore the look-elsewhere effect must be taken care of by taking into account the probability of observing a similar excess anywhere in the range \citep[see, e.g.,][]{2010EPJC...70..525G}.} or if they could simply be the
product of so-called {\it a-posteriori} statistical inference\footnote{This refers to the cases where an anomalous feature is not predicted by any
models before observing the data and is picked
arbitrarily only after looking at the data. In other words, the feature is observed because the employed statistical method is designed to detect it.}
\citep[see, e.g.,][]{bennett:2011}. Demonstrating robustness with respect
to statistics and data selection can to some extent alleviate such
criticisms. With this in mind, we study in this Letter the question of
statistical isotropy from the simplest possible point of view, namely
by computing the local variance of the CMB fluctuations over patches
of different sizes and positions on the sky, and comparing these
measurements with those obtained from isotropic simulations. Related
variance-oriented studies have been performed for example by
\citet{Bernui:2005pz,Bernui:2008ve,Lew:2008mq,Lew:2008zq,Zhao:2012hz,Rath:2013yra,Gruppuso:2013xba}.

\section{Data and Method}

We include in the following analysis the foreground-reduced co-added V (61 GHz) and W (94 GHz)
temperature sky maps from the 9-year {\it WMAP} data release, and the {\ttfamily SMICA} map from the {\it Planck} 2013
data release \citep{2013arXiv1303.5072P}; the three other {\it Planck} CMB temperature solutions
({\ttfamily Commander-Ruler}, {\ttfamily NILC} and {\ttfamily SEVEM})
give consistent results, and are omitted in the following for
brevity. To exclude pixels that are highly contaminated by diffuse
foreground emission and point sources, we adopt the {\it WMAP9} KQ85
Galactic and point source mask, with a sky coverage of $\sim75\%$, for the {\it WMAP} data, and the {\it Planck}
standardized common mask, U73, with a sky coverage of $\sim73\%$ \citep{2013arXiv1303.5072P}, for the {\it Planck} map.

\begin{figure*}[t]
\begin{center}
\subfigure{\includegraphics[width=0.43\linewidth, trim = 0 0 0 0, clip=true]{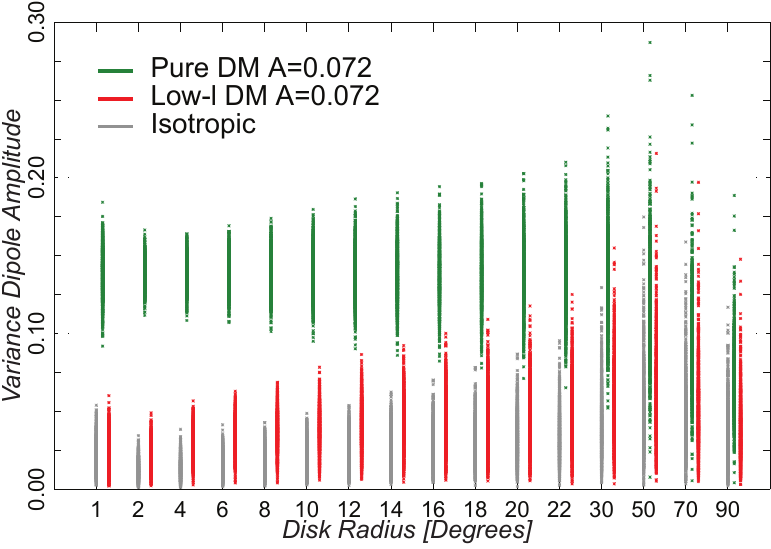}} \\
\subfigure{\includegraphics[width=0.43\linewidth, trim = 0 0 0 0, clip=true]{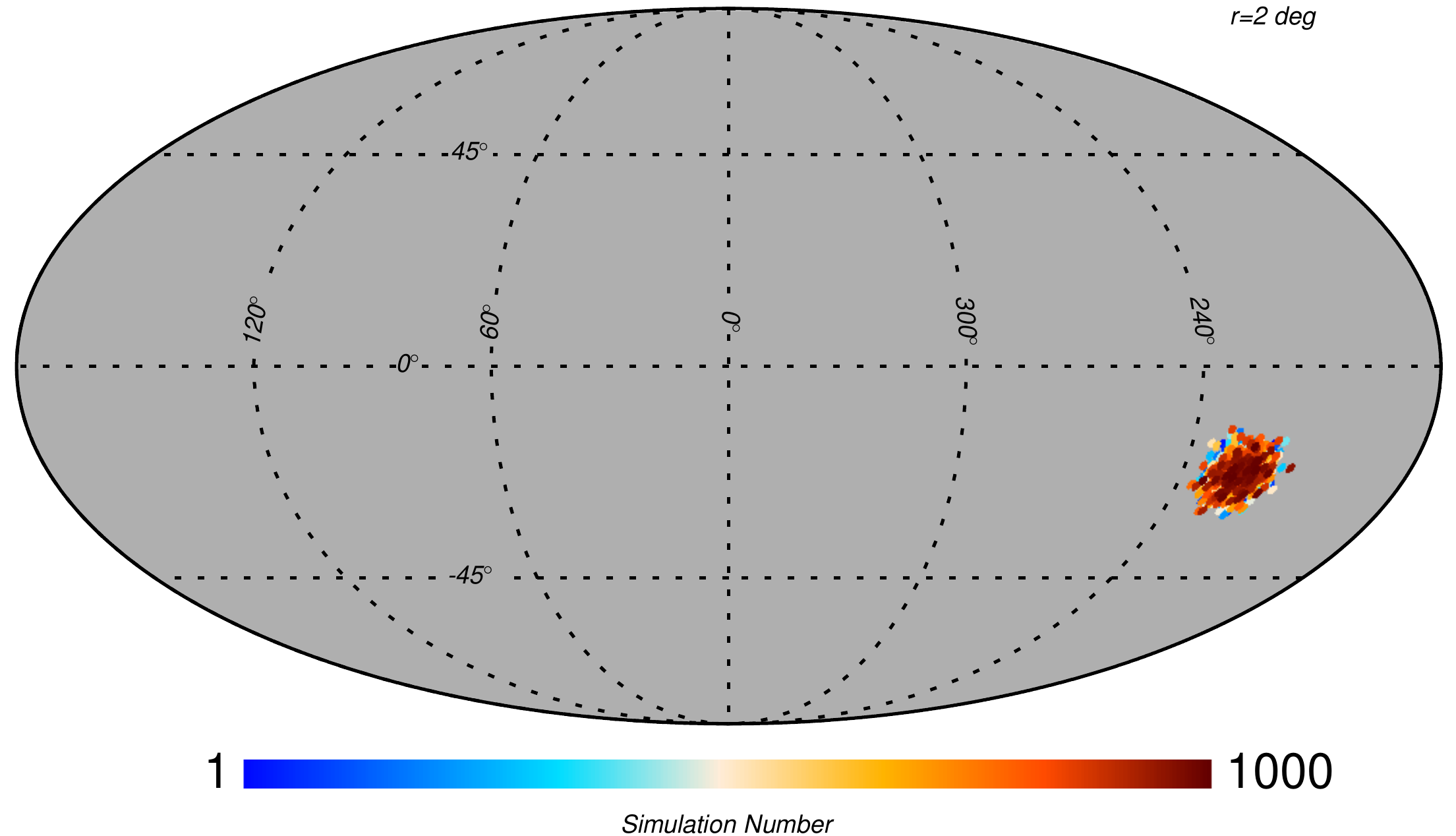}}
\subfigure{\includegraphics[width=0.43\linewidth, trim = 0 0 0 0, clip=true]{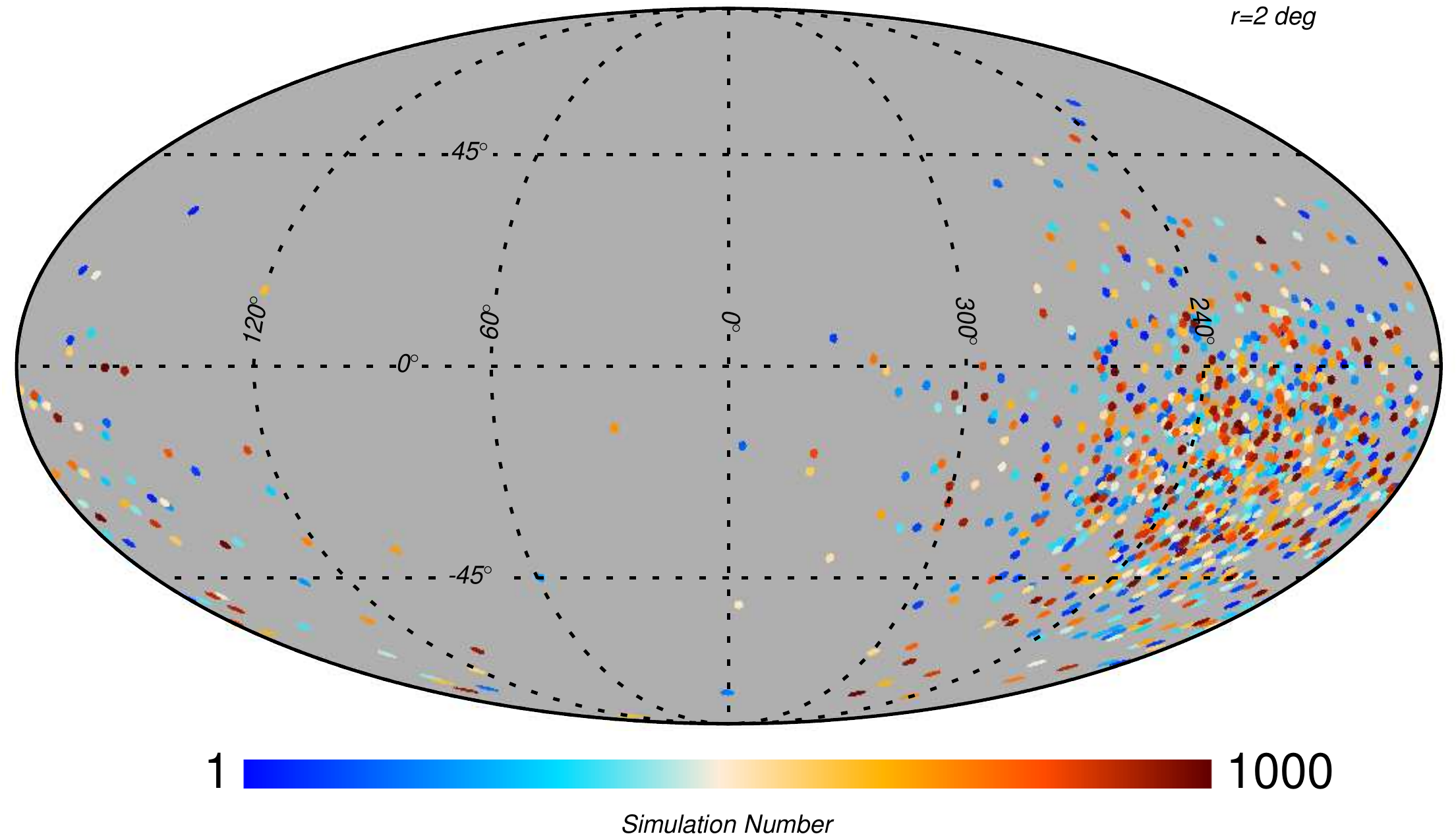}} \\
\subfigure{\includegraphics[width=0.43\linewidth, trim = 0 0 0 0, clip=true]{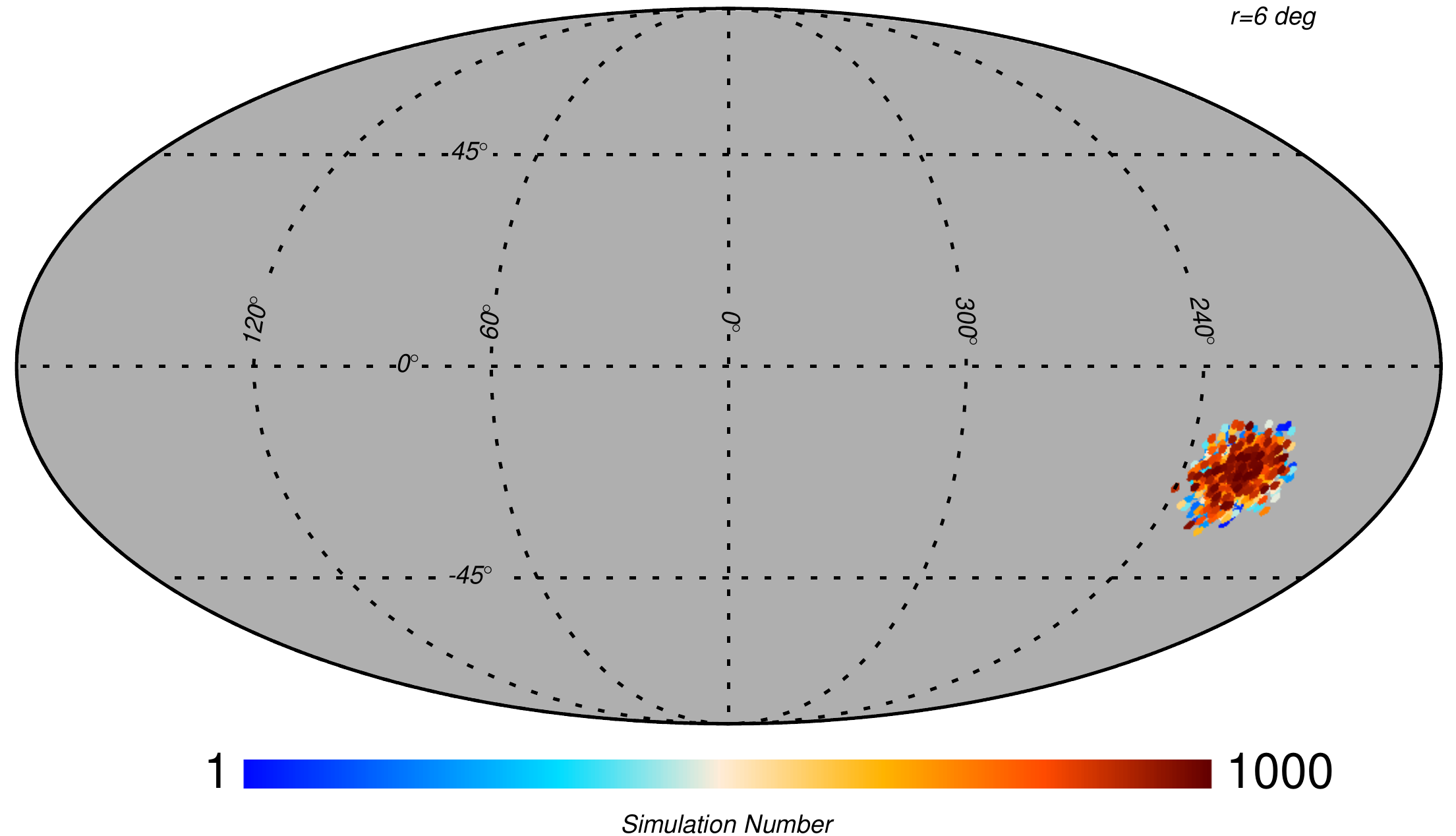}} 
\subfigure{\includegraphics[width=0.43\linewidth, trim = 0 0 0 0, clip=true]{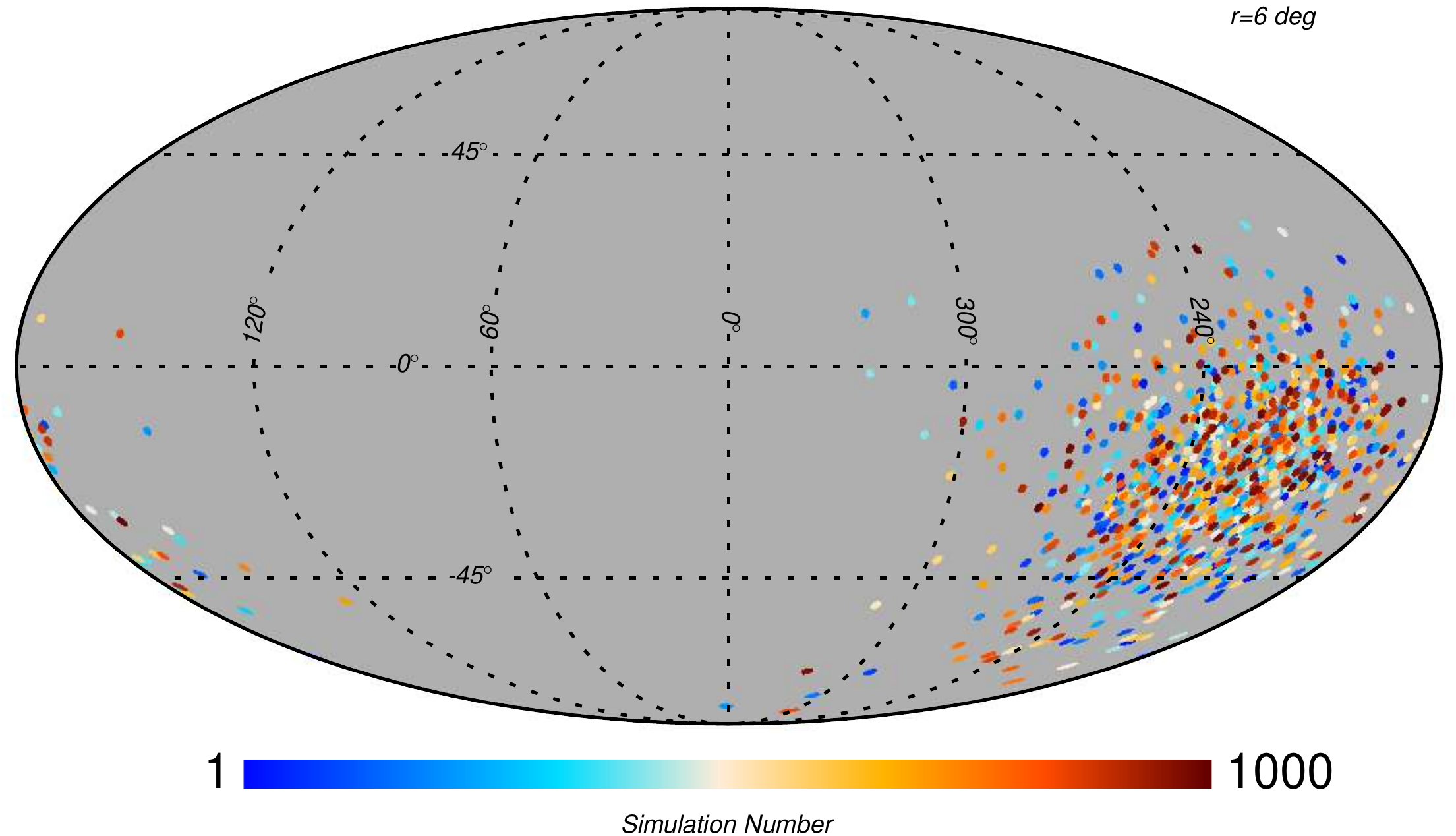}} \\
\subfigure{\includegraphics[width=0.43\linewidth, trim = 0 0 0 0, clip=true]{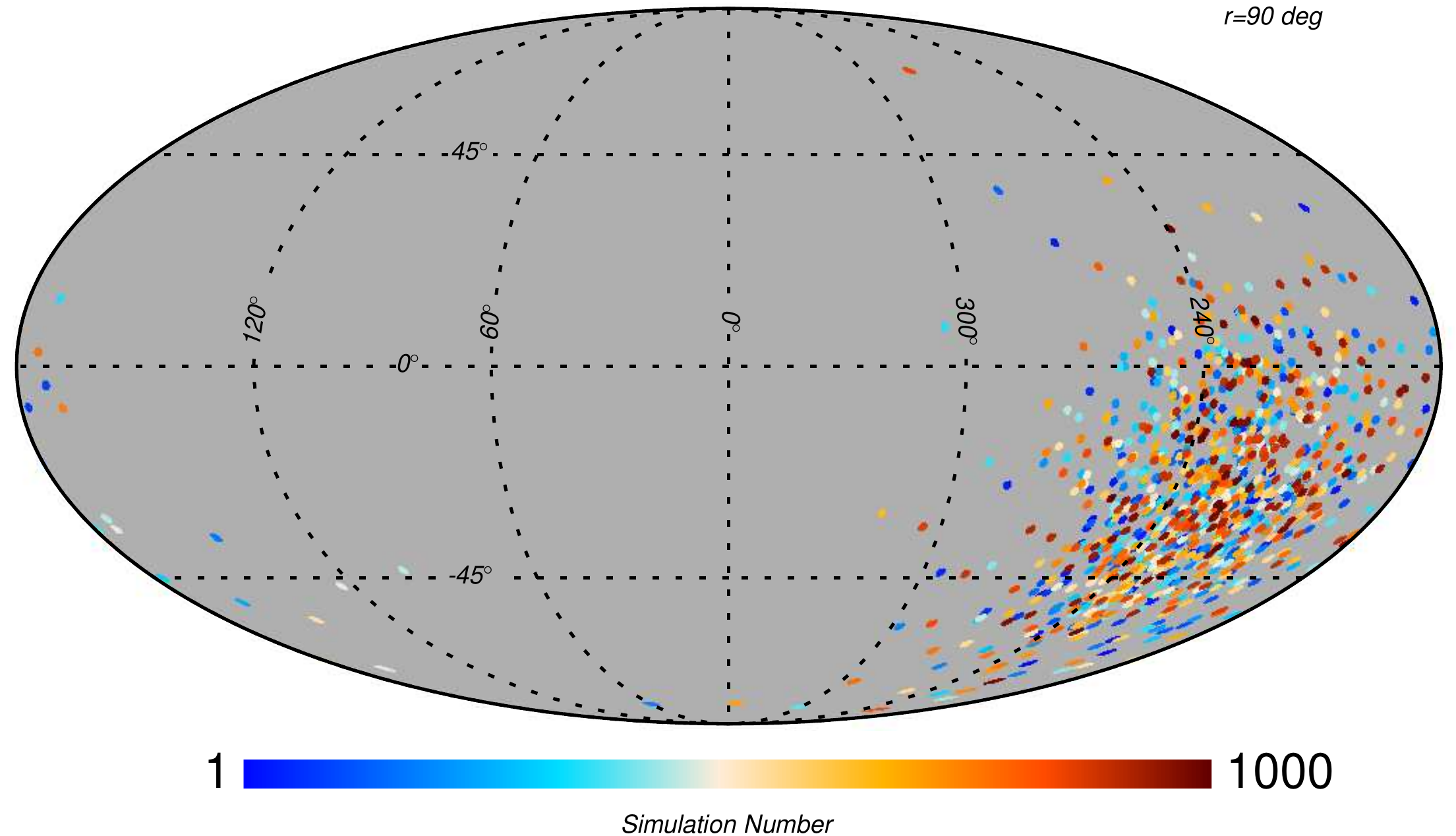}}
\subfigure{\includegraphics[width=0.43\linewidth, trim = 0 0 0 0, clip=true]{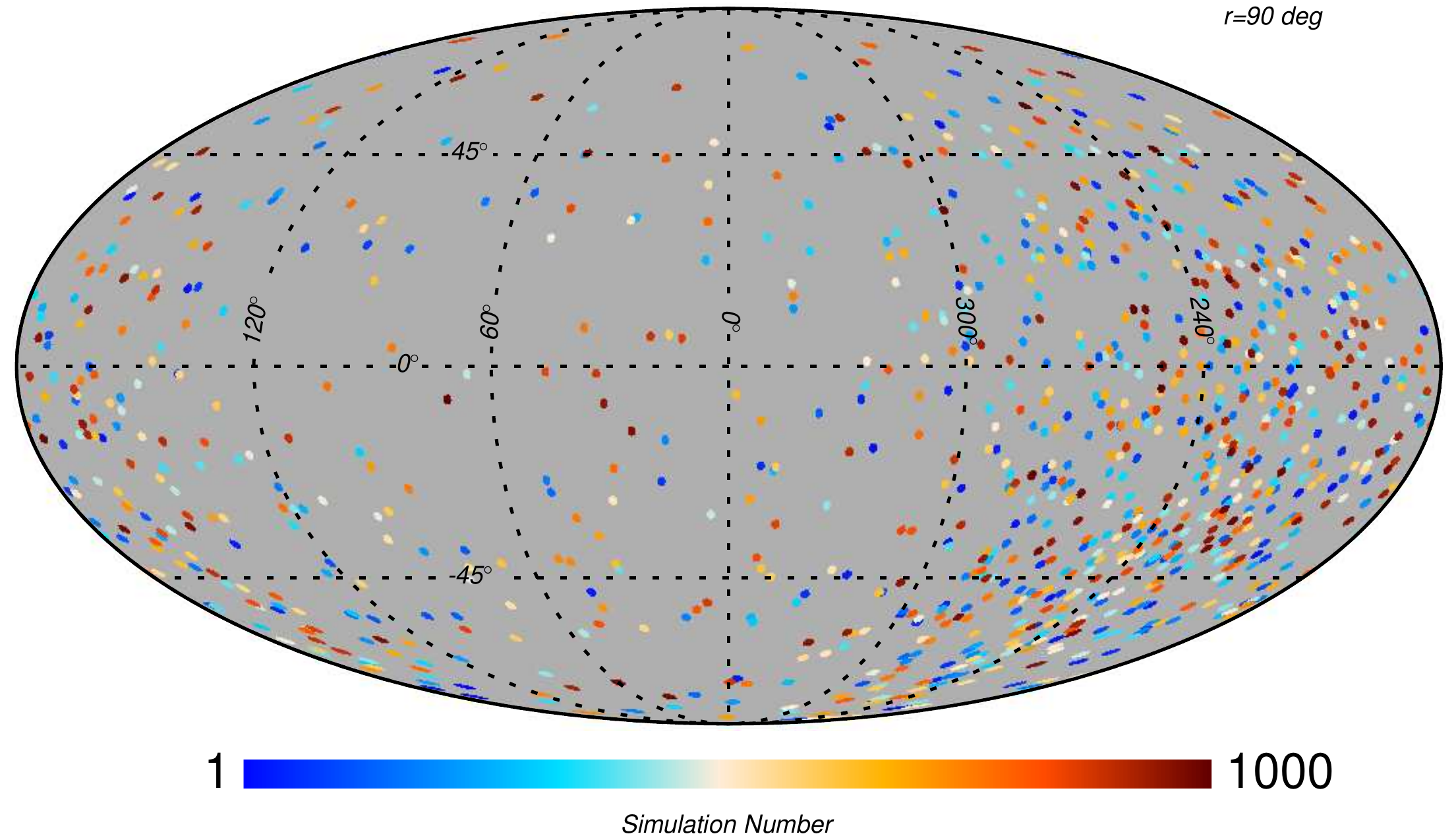}} 
\caption[aa]{\footnotesize{Top panel: local-variance dipole amplitude as a
    function of disk radius for 1000 {\it Planck} ({\ttfamily SMICA}) FFP6 isotropic simulations (gray
    points), as well as for 1000 all-scale dipolar-modulated (pure DM; green
    points) and low-multipole-only (scales larger than $5^\circ$) dipolar-modulated (low-$l$ DM; red points) simulations.  Bottom panel: dipole
    directions recovered from all-scale (first
    column) and low-multipole-only (second column) dipolar-modulated simulations
    with disks of radii $2^\circ$, $6^\circ$ and $90^\circ$.
    For all cases the input dipole amplitude and directions are $A=0.072$ and $(l,b) = (224^\circ,-22^\circ)$.}} \label{fig:calibrartion}
\end{center}
\end{figure*}

In order to assess the significance of any anisotropic signal in the
data, we resort to simulated isotropic CMB maps. For the {\it WMAP} case, we
generate 1000 CMB-plus-noise Monte Carlo simulations based on the
{\it WMAP9} best-fit $\Lambda$CDM power spectrum \citep{wmap9_2}. Noise
realizations are drawn as uncorrelated Gaussian realizations with a
spatially varying RMS distribution given by the number of observations
per pixel; the {\it WMAP} simulations do not contain lensing effects. For {\it Planck} we adopt the 1000 ``Full Focal Plane'' (FFP6)
end-to-end simulations produced by the {\it Planck} collaboration based on
the instrument performance and noise properties. The FFP6 CMB and noise maps have been propagated through the {\it Planck} pipelines with the same weighting as the data. These simulations
also incorporate lensing effects and are treated identically to the data in all steps discussed
below. The Doppler boosting effects, which have been shown to be an issue on small angular scales \citep[high multipoles;][]{2013arXiv1303.5083P}, have not been taken into account in this analysis. However, since variance is more sensitive to large angular scales (low multipoles), our results should not be significantly affected by these effects. We leave a full investigation of the Doppler boosting effects for future work. 

The analysis proceeds as follows: We consider 3072 disks (of various
sizes) centered on the pixels of a \healpix $N_{\textrm{side}}=16$
map\footnote{The results are not sensitive to the number of disks, as
  long as the entire sky is covered, and consistent results are
  obtained with, say, an $N_{\textrm{side}}=8$ grid.}
\citep{Healpix2005}. For each disk and sky map, we compute the
temperature variance including only unmasked pixels; any disk for
which more than 90\% of the area is masked is ignored completely. This
results in a low-resolution and almost full-sky map of the local
variance across the sky. To establish the expected mean and variance
of each disk, we compute the same local-variance map from the
simulated ensemble. This mean map is then subtracted from both the
observed and simulated local-variance maps, resulting in a zero-mean
variance variation map. Finally, we fit a dipole to each of these
local-variance maps using the \healpix {\ttfamily remove\_dipole} routine
using inverse variance weighting. Note that this procedure is strongly
related to the Crossing statistic described by
\citet{Shafieloo:2010xm,Shafieloo:2012jb,Shafieloo:2012yh}, which has
been applied to isotropy tests with low-redshift supernovae data
\citep{Colin:2010ds}.

\section{Analysis of anisotropic simulations}

Before we discuss our results for the real data, we assess the
sensitivity of the method by applying it to both simulated isotropic
and anisotropic CMB realizations. The anisotropic simulations have
been modulated by a dipole (Equation (\ref{eq:dipole})) with an
amplitude of $A=0.072$ and a direction of $(l, b)=(224^\circ,
-22^\circ)$, consistent with that reported for large angular scales
\citep[e.g.,][]{hoftuft2009}. Two different sets of anisotropic
simulations are generated, one for which all scales are modulated, and
another for which only scales larger than $5^\circ$ (corresponding to a $5^\circ$ smoothing) have been
modulated.

Figure \ref{fig:calibrartion} shows the resulting local-variance
dipole amplitudes (top panel) and directions (bottom panels) as a
function of disk radius for each of the 1000 FFP6 simulations, ranging
between $1^\circ$ and $90^\circ$.\footnote{The directions are shown only for anisotropic simulations; the resulting directions for isotropic simulations are uniformly distributed all over the sky and we do not show them here for brevity.} Here we see that the sensitivity of
the statistic depends significantly on the disk radius over which the
variance is computed, and at a radius of $\sim20^{\circ}$ even the
amplitude distribution for the fully modulated model starts to overlap
with the isotropic distribution. This makes intuitive sense, since the
larger the radius, the more weight is put on the larger angular
scales, and hence cosmic variance begins to dominate. For example, a
radius of $20^{\circ}$ corresponds roughly to angular features of
$\ell \approx 180^{\circ}/20^{\circ} \approx 10$. This correspondence
applies independently of the specific details of the assumed
anisotropic model, and in the following we therefore restrict our
interest to the range between $1^{\circ}$ and $20^{\circ}$.

\begin{figure*}[t]
\begin{center}
\subfigure[][]{\label{fig:dipampPlanck}\includegraphics[width=0.41\linewidth, trim = 0 -10 00 0, clip=true]{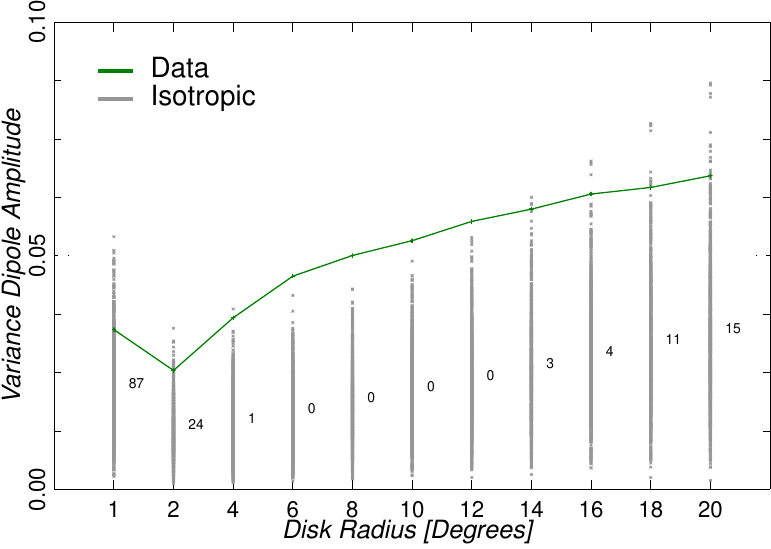}}
\subfigure[][]{\label{fig:sighistogramgaussfit}\includegraphics[width=0.45\linewidth, trim = -10 0 0 0 0, clip=true]{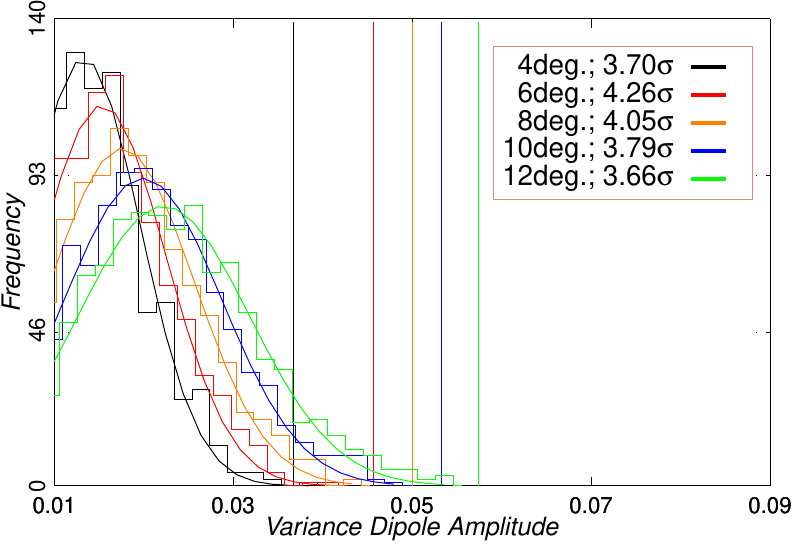}} \\
\subfigure[][]{\label{fig:varmap_6deg}\includegraphics[width=0.41\linewidth, trim = 0 -10 0 0, clip=true]{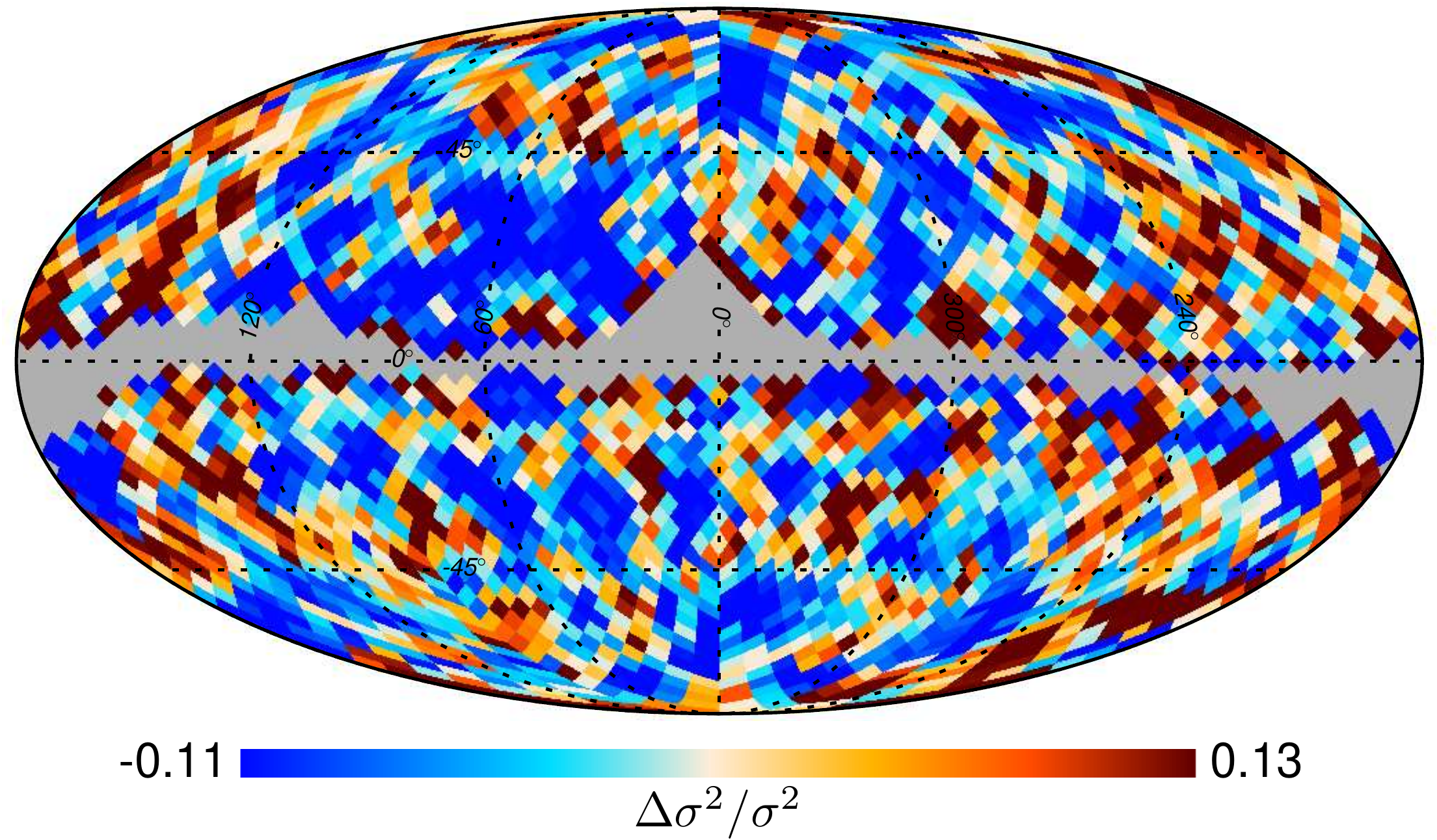}}
\subfigure[][]{\label{fig:cl_6deg}\includegraphics[width=0.44\linewidth, trim = -10 0 0 0, clip=true]{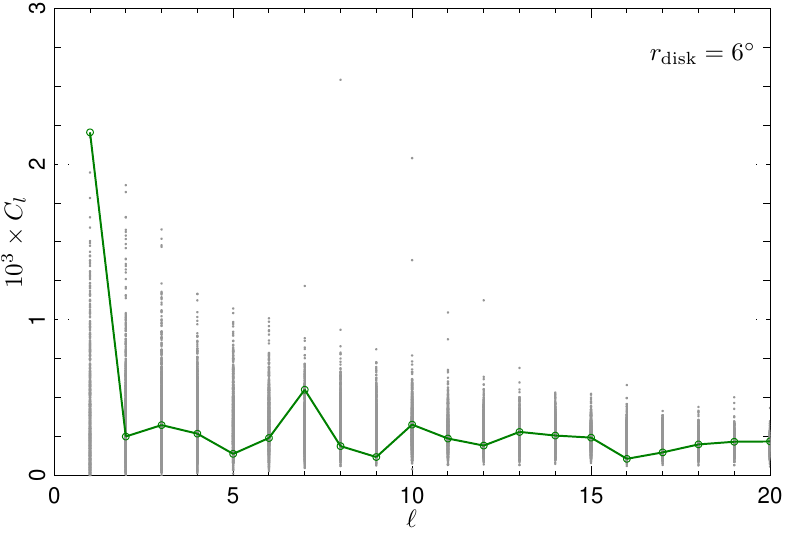}}
\caption[aa]{\footnotesize{{\bf (a)} Local-variance dipole amplitude as a
    function of disk radius for {\it Planck} ({\ttfamily SMICA}) data (in
    green) versus the 1000 isotropic FFP6 simulations (in gray). The
    labels above each scale indicate the number of simulations with
    amplitudes larger than the ones estimated from the
    data, and are located at the means of the amplitude values
from the simulations. {\bf (b)} Histograms of the local-variance dipole
    amplitudes from the 1000 FFP6 simulations for disk radii
    $4^\circ$, $6^\circ$, $8^\circ$, $10^\circ$ and $12^\circ$,
    together with the best-fit Gaussian distributions in all
    cases. Vertical lines indicate the corresponding amplitudes
    measured from the {\it Planck} data. The legend shows the rough estimates of detection significances derived from the Gaussian fits. {\bf (c)} Mean-field subtracted, local-variance map computed with $6^\circ$ disks for {\it Planck} ({\ttfamily SMICA}) data. {\bf (d)} Angular power spectrum ($C_\ell$) of the local-variance map computed with $6^\circ$ disks for the 1000 FFP6 simulations (in gray), as well as for {\it Planck} ({\ttfamily SMICA}) data (in green).}}\label{Planckresults}
\end{center}
\end{figure*}

By choosing the disk sizes this way we can also avoid the problems related to a-posteriori
statistics and look-elsewhere effects. The choice has been made before analyzing the real data, and hence, the results we will present in the next section should not be interpreted as
being a-posteriori. In addition, given that we look for the asymmetry signal only using disk sizes that are shown to be sensitive enough, our method is immune from the look-elsewhere criticism. Moreover, since the pixels on the smaller disks belong also to the larger disks, the local-variance dipole amplitudes and directions obtained using disks with different sizes are highly correlated, and this largely reduces the total number of independent statistics used. This therefore further weakens the look-elsewhere effects. One consequence of the disk correlations is that one cannot simply sum the significances from different disk sizes, nor can one use the correlations between different dipole directions to derive a statistical measure as has been done in power-asymmetry analyses \citet{hansen2009}, \citet{Axelsson:2013mva} and \citet{2013arXiv1303.5083P}. 

\begin{table*}[t]
\begin{center}
\tabletypesize{\footnotesize}
\caption{Computed Variance Asymmetry Significances with Different Disk Radii\label{tab:allsigs}}
\begin{tabular}{lllcccccccccc}
\noalign{\doubleline}
Data & Mask & $C_\ell$ Used in Simulations & $2^\circ$ & $4^\circ$ & $6^\circ$ & $8^\circ$ & $10^\circ$ & $12^\circ$ & $14^\circ$ & $16^\circ$ & $18^\circ$ & $20^\circ$ \cr
\noalign{\vskip 3pt\hrule\vskip 4pt}
{\it WMAP9} & KQ85 & {\it WMAP9} best-fit & 101 & 28 & 7 & 5 & 6 & 6 & 9 & 15 & 20 & 27 \cr
\noalign{\vskip 3pt}
{\it Planck} ({\ttfamily SMICA}) & U73 & {\it Planck} best-fit & 24 & 1 & 0 & 0 & 0 & 0 & 3 & 4 & 11 & 15 \cr
\noalign{\vskip 3pt\hrule\vskip 4pt}
{\it WMAP9} & KQ85 $+$ U73 & {\it WMAP9} best-fit & 23 & 3 & 1 & 1 & 3 & 5 & 9 & 13 & 18 & 30 \cr
\noalign{\vskip 3pt}
{\it Planck} ({\ttfamily SMICA}) & KQ85 $+$ U73 & {\it Planck} best-fit & 16 & 1 & 1 & 0 & 1 & 3 & 7 & 9 & 13 & 22 \cr
\noalign{\vskip 3pt\hrule\vskip 4pt}
{\it WMAP9} & KQ85 & {\it Planck} best-fit & 98 & 17 & 1 & 0 & 0 &  0 & 0 & 0 & 0 & 0 \cr
\noalign{\vskip 3pt\hrule\vskip 3pt}
\end{tabular}
\end{center}
\end{table*}

\section{Results}

We now apply our variance estimator to the real {\it WMAP} and {\it Planck}
data. First, Figure \ref{Planckresults}(a) shows the results for
{\it Planck}, plotted in the same format as in Figure
\ref{fig:calibrartion}, but now comparing with the 1000 FFP6 isotropic
simulations.  As seen in the plot, none of the 1000 isotropic
simulations have local-variance dipole amplitudes larger than the data
over the range $6^\circ \le r_{\small {\textrm{disk}}}\le 12^\circ$, formally
corresponding to a lower limit on the statistical significance of
$3.3\sigma$ or a $p$-value of 0.001. To give a rough estimate of an
actual significance, and not only a lower limit, we plot in Figure
\ref{Planckresults}(b) histograms of the variance dipole
amplitudes for the FFP6 simulations at the disk radii with the highest
detection significances, and fit a Gaussian in each case. Employing
these extrapolations, we derive significances of $\sim4\sigma$ in each
of these cases. However, we emphasize that these numbers only serve as
a rough guide, as the distributions do have significant non-Gaussian
tails; an extended ensemble of simulations is certainly preferable
over this approximation.

In order to see how a typical local-variance map, for a high-significance detection of asymmetry, looks like, we show in Figure \ref{Planckresults}(c) the mean-field subtracted map for $6^{\circ}$ disks. Figure \ref{Planckresults}(d) shows the angular power spectrum of the same map as a function of multipoles. This clearly shows that the dipole component is the dominant mode in the local-variance map. Figure \ref{Planckresults}(d) also indicates that the data is very consistent with isotropic simulations at all multipoles except for the dipole, which is anomalously large.

Similar results derived from the {\it WMAP9} observations show qualitatively
the same trend, but differ somewhat in terms of final
significances (see Table \ref{tab:allsigs}). Specifically, a maximum asymmetry is seen between
$6^{\circ}$ and $12^{\circ}$, with $\sim5$ out of 1000 isotropic
simulations exhibiting a larger variance dipole amplitude, for a $p$-value of $\sim0.005$ and a statistical significance of $\sim 2.9 \sigma$.

\begin{figure*}[t]
\begin{center}
\label{alldirections}\includegraphics[width=0.9\linewidth, trim = 0 0 0 0, clip=true]{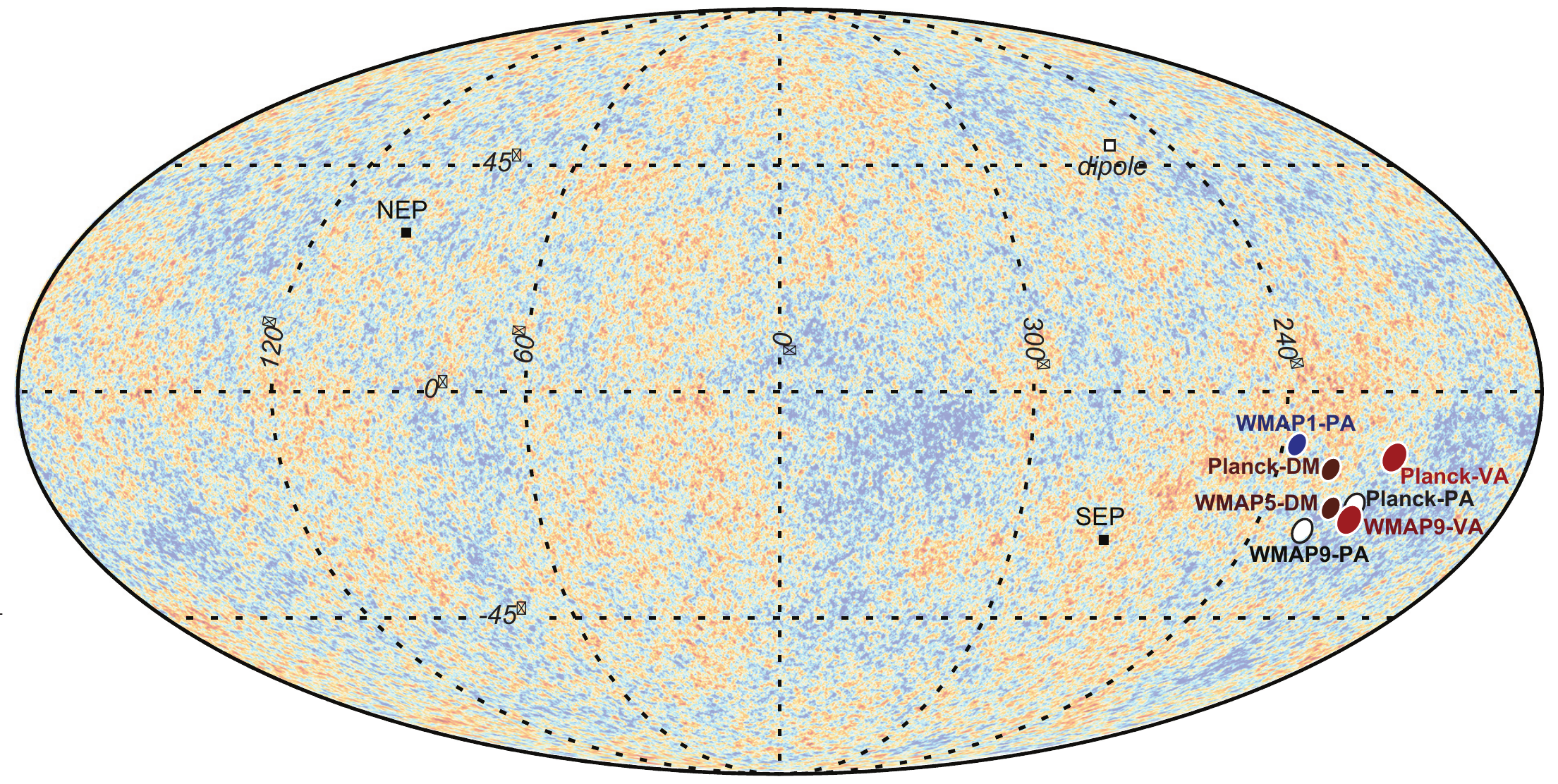} \\
\caption[aa]{\footnotesize{Asymmetry directions found in this work by analyzing the local variance of the {\it WMAP} 9-year and {\it Planck} 2013 data [denoted by {\it WMAP9}-VA and {\it Planck}-VA], as well as the directions found previously from the latest likelihood analyses of the dipole modulation model [denoted by {\it WMAP5}-DM \citep{hoftuft2009} and {\it Planck}-DM \citep{2013arXiv1303.5083P}] and the local-power spectrum analyses [denoted by {\it WMAP1}-PA
\citep{eriksen2004}, {\it WMAP9}-PA
\citep{Axelsson:2013mva} and {\it Planck}-PA \citep{2013arXiv1303.5083P}] for the {\it WMAP} and {\it Planck} data. The background map is the CMB sky observed by {\it Planck} ({\ttfamily SMICA}). VA, DM and PA stand for variance asymmetry, dipole modulation and power asymmetry, respectively.}} \label{fig:alldirections}
\end{center}
\end{figure*}

The preferred directions obtained in the present Letter (for
$8^{\circ}$ disks) are listed in Table \ref{tab:alldirections},
together with a number of similar results obtained in previous papers,
and summarized visually in Figure \ref{fig:alldirections}.\footnote{Note that no dipole modulation results have been published for the 9-year
{\it WMAP} temperature sky maps to date. However, given that the 9-year {\it WMAP} sky
maps are virtually indistinguishable from the 5-year sky maps on angular
scales larger than $5^{\circ}$ relative to cosmic variance, we expect the
dipole modulation results to be very close to those reported by \citet{hoftuft2009} for the 5-year {\it WMAP} data.} Clearly,
the preferred directions derived by very different algorithms and data
combinations are all in good qualitative agreement.

\paragraph{{\it Ecliptic variance asymmetry:}}

Before we end this section, we present the results of a different, but related study, using variance as the statistic, which we have performed in addition to the main analysis of this Letter. In \citet{2013arXiv1303.5083P} it is reported that the variances of the CMB fluctuations computed on the northern ecliptic and Galactic hemispheres are significantly smaller compared to the corresponding southern ones. We repeated the same analysis here, for the ecliptic hemispheres, and obtained similar results. A potential criticism of this study is the fact that it ignores look-elsewhere effects. This can be dealt with by using a test statistic, based on the differences in variances between different hemispheres, that involves a ranking procedure. Specifically, we first compute the difference in variances for each pair of opposite hemispheres for the data. By sorting the obtained values we assign a rank to for example the northern-southern ecliptic hemispheres. We then compare this value to the values with the same rank obtained from repeating the same procedure to all isotropic simulations, and derive a $p$-value. The $p$-value we obtain this way for the {\it Planck} data and FFP6 simulations shows that the variance difference along the ecliptic pole for the data is not significantly different from that of the isotropic simulations. However, performing the same procedure on the variance values for hemispheres (and not the variance differences) indicates that the variance from the northern ecliptic hemisphere for the data is still significantly low with a $p$-value of $4/1000$. Using ratios of variances instead of differences in variances has no impact on our results.

\begin{table}[]
\begin{center}
\tabletypesize{\footnotesize}
\caption{Asymmetry Directions\label{tab:alldirections}}
\scalebox{.87}[1.0]{
\begin{tabular}{l l l l}
\noalign{\doubleline}
Map & ($l$,$b$) $[\deg]$ & Significance & Reference \cr
\noalign{\vskip 2pt}
 &  & or $p$-value & \cr
\noalign{\vskip 3pt\hrule\vskip 3pt}
{\it Planck}-VA\tablenote{VA, DM and PA stand for variance asymmetry, dipole modulation and power asymmetry, respectively.} & (212, -13) & $0/1000$ & Present work \cr
\noalign{\vskip 4pt}
{\it WMAP9}-VA  & (219, -24) & $5/1000$ & Present work \cr
\noalign{\vskip 3pt\hrule\vskip 3pt}
{\it Planck}-DM   & (227, -15) & $3.5\sigma$ & \citet{2013arXiv1303.5083P} \cr
\noalign{\vskip 4pt}
{\it WMAP5}-DM & (224, -22) & $3.3\sigma$ & \citet{hoftuft2009} \cr
\noalign{\vskip 4pt}
{\it Planck}-PA   & (218, -21)  & $0/500$ & \citet{2013arXiv1303.5083P} \cr
\noalign{\vskip 4pt}
{\it WMAP9}-PA   & (227, -27) & $7/10000$ & \citet{Axelsson:2013mva} \cr
\noalign{\vskip 3pt\hrule\vskip 3pt}
\end{tabular}}
\end{center}
\end{table}

\section{Summary}

We have applied a simple local-variance estimator to the latest {\it WMAP}
and {\it Planck} data, performing a frequentist test of global statistical
isotropy. For the {\it Planck} data, we find that the local variance
exhibits dipolar-like spatial variations that are statistically
significant (at least) at the $\sim3\sigma-3.5\sigma$ level on scales between $\sim4^{\circ}$
and $14^{\circ}$ by this measurement. For {\it WMAP}, we find a statistical
significance of $\sim2.9\sigma$, and a direction fully consistent with
that derived from {\it Planck}. 

The results obtained here are in good qualitative agreement with
earlier results, for example using direct likelihood fits or bipolar
harmonics \citep{hoftuft2009,2013arXiv1303.5083P}, that indicate a
$\gtrsim3\sigma$ dipole-modulation-like effect on large angular
scales, but with an amplitude that is decreasing with angular
scale. In the present approach, this is seen by the fact that the
statistical significance decreases for the smallest disk radii of
$1^{\circ}-2^{\circ}$ scales; for a pure dipole modulation extending through
all multipoles this stays constant.

The slight difference between {\it Planck} and {\it WMAP} can be mainly explained by the different masks that have been used in the two cases. In order to study the effects of the masks on the results, we repeated the analysis for both the {\it WMAP} and {\it Planck} maps using a unified mask, i.e. the combination of the {\it Planck} U73 and {\it WMAP9} KQ85 masks. We obtained very similar results in these cases (see Table \ref{tab:allsigs}). The remaining differences can be at least partially explained by the different noise levels of the two
experiments. No smoothing is applied to either data set in this
analysis, and the variance therefore receives a significant
contribution from the pixel-scale noise, which is substantially larger
for {\it WMAP} than for {\it Planck}, decreasing the effective sensitivity of the
estimator.

One should note however that choosing the right theoretical angular power spectrum for the isotropic simulations of the CMB maps is crucial and can also impact the obtained statistical significances, and might therefore provide another explanation for the differences in the significances computed from {\it WMAP} and {\it Planck}. It is known that \citep{2013arXiv1303.5075P} there is a $1\%-2\%$ mismatch between the power spectra computed from the two data sets and this can potentially explain some of the discrepancy that we see here. We have tested this effect by using the {\it Planck} best-fit power spectrum \citep{2013arXiv1303.5075P} to generate the simulations for {\it WMAP}. We observe that enforcing the {\it Planck} spectrum on the {\it WMAP} data makes the {\it WMAP} results comparable with the {\it Planck} ones. The significances for {\it WMAP} are now 0/1000 over disk radii of $8^\circ-22^\circ$ (see Table \ref{tab:allsigs}), while the dipole directions do not change. This suggests, and remains to be investigated, that after resolving the tension between the {\it Planck} and {\it WMAP} power spectra the results of our analysis for the two experiments will agree better.

In this Letter we have focused on dipolar variations in the local-variance map, but this is clearly easily generalizable to higher-order
modes. This will be considered in future work. For now, we note that
the main advantages of the method are its conceptual and
implementational simplicity, its directly intuitive interpretation,
and, by virtue of being defined in pixel space, a useful
complementarity to other typically harmonic-based methods. The fact
that different statistical techniques, with different properties and
sensitivities, result in very similar conclusions does weaken the
suggestion that the effect is simply the product of a-posteriori
statistics. 

\begin{acknowledgments}
We thank Claudio Llinares, Eamon M. Scullion and Amir Hajian for
helpful discussions. Y.A. and H.K.E. acknowledge
support through the ERC Starting Grant StG2010-257080. Y.F. is supported by ERC Grant 277742 Pascal. A.S. thanks the
Korean Ministry of Education, Science and Technology (MEST),
Gyeongsangbuk-Do and Pohang City for the support of the Independent
Junior Research Groups at the Asia Pacific Center for Theoretical
Physics (APCTP). F.K.H. acknowledges OYI grant from the Norwegian research council. We acknowledge the use of resources from the
Norwegian national super-computing facilities, NOTUR. Maps and results
have been derived using the \healpix (http://healpix.jpl.nasa.gov)
software package developed by \cite{Healpix2005}.
\end{acknowledgments}

\end{document}